\pdfoutput=1

\documentclass[twocolumn,showpacs,amsmath,amssymb,superscriptaddress,prb]{revtex4-1}

\usepackage{graphicx}
\usepackage{epstopdf}
\usepackage{dcolumn}
\usepackage{bm}

\begin{document}


\title{Vacuum Compatibility of 3D-Printed Materials}



\author{A. P. Povilus}
\affiliation{Department of Physics, University of California at Berkeley, Berkeley, CA 94720-7300, USA}
\email[]{povilus@berkeley.edu}
\author{C. J. Wurden}
\affiliation{Department of Physics, University of California at Berkeley, Berkeley, CA 94720-7300, USA}
\author{Z. Vendeiro}
\affiliation{Department of Physics, University of California at Berkeley, Berkeley, CA 94720-7300, USA}
\author{M. Baquero-Ruiz}
\affiliation{Department of Physics, University of California at Berkeley, Berkeley, CA 94720-7300, USA}
\author{J. Fajans}
\affiliation{Department of Physics, University of California at Berkeley, Berkeley, CA 94720-7300, USA}


\date{\today}

\begin{abstract}
The fabrication fidelity and vacuum properties are tested for currently available 3D-printed materials including polyamide, glass, acrylic, and sterling silver.  The silver was the only material found to be suitable to ultrahigh vacuum environments due to outgassing and sublimation observed in other materials.
\end{abstract}


\maketitle

\section{Introduction}

Additive manufacturing, or 3D printing, of parts has many potential advantages over traditional machining for construction of experimental apparati.  \cite{SLSFab1}   Material and machining costs of parts fabricated using 3D-printing methods are generally lower since there is little waste in the process and complicated shapes become much easier to produce.  Printing also gives the ability to generate geometries that would have only been previously possible with either welding or an expensive casting process.  Since printing is additive, it is possible that there are small voids in the material that would trap gases that would later vent slowly into the vacuum, making the part unsuitable for ultrahigh vacuum environments. \cite{SLSFab2}  It is of particular interest how well these materials perform in a vacuum environment since they can drastically reduce the time and cost to build complicated, specialized geometries, such as resonant electrode strucutres, for use in cold atom traps and plasma devices.

\section{Printed Part Description}

\begin{figure}
\includegraphics{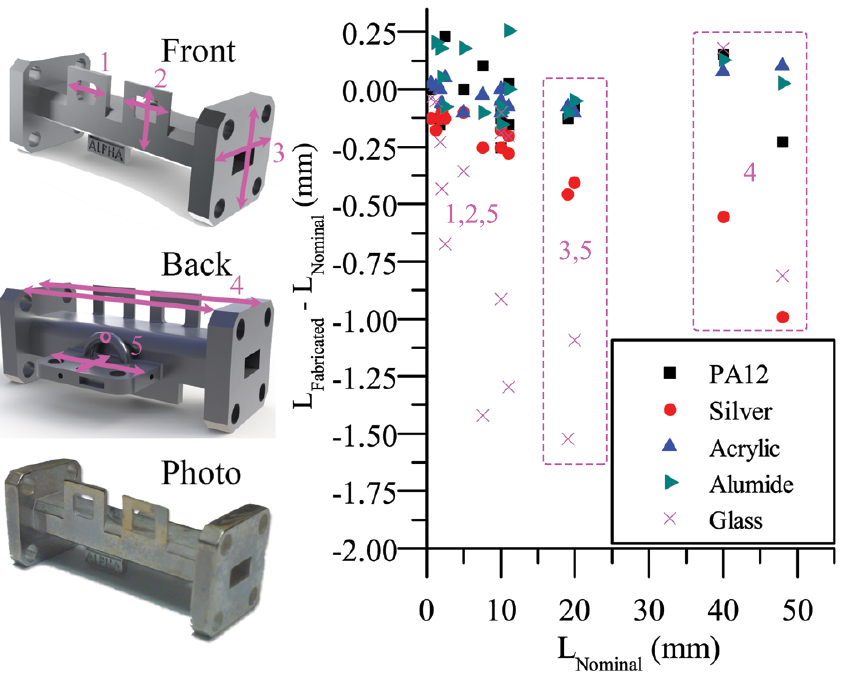}
\caption{(Color online) CAD renditions (\emph{ top left}) of a modified waveguide used as an sample 3D printed part compared a photograph (\emph {bottom left}) of the actual waveguide printed in sterling silver.    The lengths of particular features of this waveguide, highlighted in the CAD renders, are compared to printed versions of waveguide in different materials.  The difference of the measured length of each waveguide to the nominal length in the CAD submission is shown for these features in the graph. (\emph{right})  The waveguide has dimensions $48\rm{mm} \times 20\rm{mm} \times 19\rm{mm}$.  Measurements were made to $0.01 \rm{mm}$ accuracy and precision.  \label{fig:F1}}.  
\end{figure}

In order to test the dimensional tolerances and outgassing properties of printed materials, a CAD design for a modified waveguide was printed by Shapeways in a variety of different materials using printers developed by EOS GmbH.  \cite{MaterialDatasheets}  These printers use a technique called selective laser sintering (SLS) to fuse powdered materials into parts. \cite{SLSPatent}   While the waveguides were ordered from a specific printing service, they are thought to be representative of 3D printed materials by examination of material datasheets from similar sintering printer services that use the same methods.   The waveguide was modified to include extra exterior tabs to measure dimensional tolerances of the printing process.  Printed materials included glass, acrylic, PA12 polyamide, alumide (polyamide/aluminum powder), and sterling silver ($92.5\%$ Ag). Although stainless steel was also available, we did not use it in our system since it was found to be a highly-magnetic alloy incompatable with magnetic confinement experiments of interest to us.   It is important to note that sterling silver printing is a cast part where the mold has been 3D printed, so it is not expected to have trapped gases that the other materials may have.

Anomalies in the printed parts were noted for these waveguides. Glass was found to be unsuitable for fine detail, as it warped significantly such that surfaces designed to be flat became noticeably concave ($> 0.5 \rm{mm}$ deflection), even with sufficient support during printing.  Common failure modes in others materials included minor distortions in thickness or breaks, especially in a thin tab ($0.65 \rm{mm}$) designed to test fabrication of small features.  At the present, fractures and poor resolution in thin features is a known issue with printing fine-detailed objects fabricated by a laser sintering process.  \cite{SLSGuide}  The thin acrylic tab broke off at the base during shipping and the silver tab, while solidly attached, had a small fracture running through the material.  The adjacent, thicker tab ($1.29 \rm{mm}$) did not have these defects in either the silver or acrylic.  The printed silver and acrylic were found to have excellent machinability, so some small features may still be possible by traditional machining methods after printing an oversized piece.  

The accuracy and precision of printed part dimensions was measured by comparing the lengths of features on the waveguide to the nominal specifications in the CAD model.  The difference between the nominal and measured lengths are compared in Fig.~\ref{fig:F1}.  Note that the silver part was found to be scaled down by $ \sim 1.6 \% $  relative to the CAD model; this was confirmed by the printing service to be a miscalibration of the scaling in the printer.  In order to compensate for this, we measured accuracy as the variance from the residual sum of squares assuming a linear offset.  The accuracy was found to be $\pm 0.072 \rm{mm}$ and $\pm 0.138 \rm{mm}$ for silver and polyamide respectively.

\section{Vacuum Properties}

\begin{figure}[t]
\includegraphics{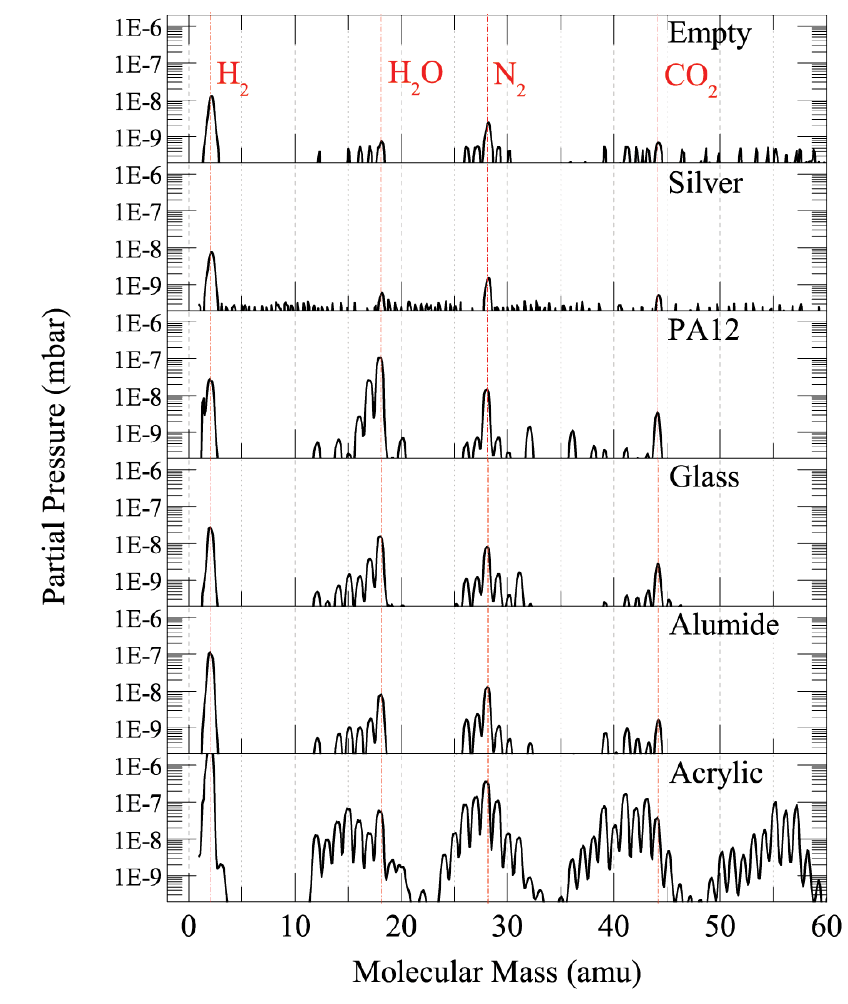}
\caption{Residual gas analysis of materials in a vacuum system 24 hours after cleaning and bakeout processes.  The top curve is the residual gas of an empty chamber.   \label{fig:F2}}
\end{figure}

Before measuring the outgassing behavior of the materials, the waveguides were chemically cleaned and baked in vacuum to remove any oils or dust.  Using CERN cleaning guidelines,\cite{CleaningMethods} adapted as needed to suit the variety of materials, the following general procedure was used to clean parts:

\begin{enumerate}
\item Ultrasonically clean part in deionized water with mild detergent
\item Rinse with deionized water
\item Degrease using acetone if material is acetone resistant
\item Ultrasonically clean with ethanol heated at 40C to dissolve excess organics
\item Evaporate excess ethanol by spraying with dry nitrogen
\item Insert into vacuum chamber and pump down to $10^{-6} \, \rm{mbar}$ or better pressure
\item Bake chamber at 24 hours at minimum of either $120 ^{\circ} \rm{C}$ or maximum recommended temperature of the material.  Ramp temperatures up and down at a maximum of 5C/hr.
\end{enumerate}

The baking and residual gas analysis was performed in the same vacuum chamber.  The 25L, room-temperature vacuum chamber was brought to ultrahigh vacuum by a turbopump backed by a dry rotary roughing pump and maintained using an 150L/s ion pump.  The empty chamber was able to be pumped to an ultimate pressure of $2.0 \cdot 10^{-9} \, \rm{mbar}$.  With the gas analyzer filament on, the empty chamber pressure rose to  $1.2 \cdot 10^{-8} \, \rm{mbar}$,  where nearly $90 \%$ of the total pressure is due to molecular hydrogen.  This is a typical behavior found in hot-filament vacuum gauges such as the gas analyzer due to adsorption processes. \cite{readhead87}

The residual gas analysis was performed using a SRS RGA100 mass spectrometer 24 hours after the bakeout was completed.  The spectrometer filament was placed out of line of sight of the test piece to avoid excess heating;  results are shown in Fig. \ref{fig:F2}.  The silver piece had outgassing rates below $1 \times 10^{-10} \rm{ mbar\, L/cm^2 s}$, the detection limit of our analyzer.  The polyamide and alumide had outgassing rates of $\sim   3 \times 10^{-8} - 4\times 10^{-7}  \rm{ mbar\, L/cm^2 s}$ following baking,  comparable to teflon and viton materials.\cite{Peacock80}  The residual gases present were atmospheric, suggesting that air was trapped in the material.  Alumide, although part polyamide, appears to have adsorpted less water from the cleaning process.  To avoid sublimation of polyamide material itself, the polyamide based materials must be baked at a low temperature ($65 ^{\circ} \rm{C}$).  When attempts were made to bake at $100 ^{\circ} \rm{C}$, the vacuum became very poor ($\sim 10^{-6} \rm{mbar}$) and a residue was found on the vacuum chamber walls near the piece.  Glass had outgassing performance comparable with the polyamide materials, but could be baked at higher temperatures.

Acrylic was not able to be pumped to ultrahigh vacuum pressure levels.  The mass spectrum for acrylic appears to have a hydrocarbon chain contamination, shown as the $13 \rm{amu}
$ repeating feature in the mass spectrum, leading to the conclusion the acrylic material was sublimating into vacuum, even at low ($55^{\circ}\rm{C}$) baking temperatures.  

\section{Conclusion}

 Of the tested materials, only sterling silver was found to be accurately printed and UHV-compatible; polyamide-based materials could be used sparingly in vacuum similar to teflon and viton.  Glass and acrylic are not recommended since glass tended to deform during fabrication, and acrylic had very poor bakeout performance.

\end{document}